\documentclass[epj]{svjour}
\pdfoutput=1

\usepackage{hyperref}
\hypersetup{
    colorlinks=true,
    urlcolor=blue,
    citecolor=black,
    linkcolor=black
}
\usepackage{amsfonts,amsmath,amssymb}
\usepackage{graphicx}
\usepackage{color}
\usepackage[square,sort,comma,numbers]{natbib}

\begin{document}

\newcommand{\apjl}{Astrophys. J. Lett. }
\newcommand{\apjs}{Astrophys. J. Suppl. Ser. }
\newcommand{\aap}{Astron. \& Astrophys. }
\newcommand{\aj}{Astron. J .}
\newcommand{\apj}{Astrophys J. }
\newcommand{\nat}{Nature }
\newcommand{\prd}{Phys. Rev D. }
\newcommand{\araa}{Ann. Rev. Astron. Astrophys. } 
\newcommand{\mnras}{Mon. Not. R. Astron. Soc. }
\newcommand{\jcap}{JCAP }
\newcommand{\pasj}{PASJ }
\newcommand{\pasa}{Pub. Astro. Soc. Aust. }

\title{Galactic Shapiro Delay to the Crab Pulsar and limit on Weak Equivalence Principle Violation}
\author{Shantanu  Desai \inst{1} \and Emre Kahya\inst{2} }

\institute{$^{1}$Department of Physics, Indian Institute of Technology, Hyderabad, Telangana-502285, India \\ $^{2}$Department of Physics, Istanbul Technical University, Maslak 34469 Istanbul, Turkey}

%

\authorrunning{S. Desai and E. Kahya}
\titlerunning{Shapiro delay to Crab Pulsar}
\abstract{
We calculate the total galactic Shapiro delay to the Crab pulsar by including the contributions from the  dark  matter as well as baryonic matter along the line of sight. The total delay due to dark matter potential is about 3.4 days.
For baryonic matter, we included the contributions from both the bulge and the disk, which are  approximately 0.12 and 0.32
days respectively. The total delay from all the matter distribution is  therefore 3.84 days. We also calculate the limit 
on violations of Weak equivalence principle by using observations of ``nano-shot'' giant pulses from the Crab pulsar with time-delay $<0.4$~ns, as well as using time differences between radio  and optical photons observed from this pulsar. Using the former, we obtain a limit on violation of Weak equivalence principle in terms of the PPN parameter $\Delta \gamma < 2.41\times 10^{-15}$. From the time-difference between simultaneous optical and radio observations, we get $\Delta \gamma < 1.54\times 10^{-9}$. We also point out differences in our calculation of Shapiro delay and that from two recent  papers~\cite{BZhang,YZhang}, which used the same observations to obtain a corresponding limit on $\Delta \gamma$.
}

\maketitle

\section{Introduction}

In 1964, I. Shapiro~\cite{Shapiro} pointed out that the  round-trip time of an electromagnetic pulse to the inner planets of our solar system experiences a delay  due to the  non-zero gravitational potential of the Sun. This delay  is referred to  in the literature as    ``Shapiro delay'' and has been measured precisely in the solar system for more than five decades, enabling very stringent tests of  general relativity (GR)~\cite{Will} and also  an astrophysical probe to measure neutron star masses in binary systems~\cite{Demorest}. Following the detection of neutrinos from SN~1987A~\cite{IMB,Kamioka}, it was pointed out  that  the neutrinos also encountered  a Shapiro delay of about 1-6 months due to the gravitational potential of the intervening matter along the line of sight~\cite{Longo,Krauss}. From the first measured GW signal GW150914~\cite{LIGO}, one can deduce that the Shapiro delay for GWs is frequency independent~\cite{Kahya16,Gao}. The most recent electromagnetic observations seen in association with  GW170817, show that 
gravitational waves also experience the same Shapiro delay as photons to about $\mathcal{O} (10^{-8})$~\cite{LVCFermi,Meszaros17,Boran,Murase,Fan}.

Wei et al ~\cite{Wei} (following earlier suggestions  in ~\cite{Sivaram}),  have  pointed out that  one can use the relative time difference between different astrophysics messengers seen across a broad swath of frequencies, to constrain energy-dependent  or inter-messenger Shapiro delay violations, enabling us  to set a stringent limit on any violations of weak equivalence principle.~\footnote{We note that some of these works eg.~\cite{Wei,Longo} refer to these  as tests   of Einstein's equivalence principle (EEP). However, EEP entails three different assumptions, among  which WEP is one of them~\cite{Will}. In other words, constancy of line of sight Shapiro delay is  mainly a test of whether the different messengers propagate on the same null geodesics.}
WEP violation is predicted in many quantum gravity models~\cite{damour} and the equality of Shapiro delays across the electromagnetic spectrum can be used to test for such violations. The violation of WEP is usually parameterized in terms of the post-Newtonian parameter $\Delta \gamma$~\cite{Will}. This technique has been applied to EM observations from a wide variety of extra-galactic astrophysical objects such as FRBs~\cite{Wei,Kaplan}, blazars~\cite{blazar}, GRBs~\cite{grb,Wei2}, etc. A complete summary of limits on WEP principle can be found in Wu et al~\cite{Wu17}. Most recently, two  independent groups have using the Crab pulsar to obtain limit on violations of WEP~\citep{BZhang,YZhang}. Yang and Zhang~\cite{BZhang} (hereafter Y16) have used ``nano-shot'' giant pulses from the Crab pulsar with time-delay between different energies of about 0.4 nanoseconds~\cite{nanoshot} to set the most stringent limit on the violation of equivalence principle of $\Delta \gamma < (0.6-1.8) \times 10^{-15}$. The Crab
pulsar has been  simultaneously timed at radio, optical, X-Ray, and $\gamma$-ray wavelengths. Zhang and Gong~\cite{YZhang} (hereafter Z16) have used the arrival time difference between various combinations of the above observations to obtain a limit on $\Delta \gamma <(2.63-4.01) \times 10^{-9}$. In this work, we calculate the total Shapiro delay to the Crab pulsar (by incorporating both the dark matter and baryonic contribution) using the same method used to calculate Shapiro delay to a variety of galactic sources in our previous works~\cite{Desai,Kahya10,Desai15}. From the calculated Shapiro delay, we then
obtain constraints on WEP using the same observations as in  Y16 and Z16.

\section{Estimated Shapiro delay to Crab Pulsar}

\begin{figure}
\centering
\includegraphics[width=0.5\textwidth]{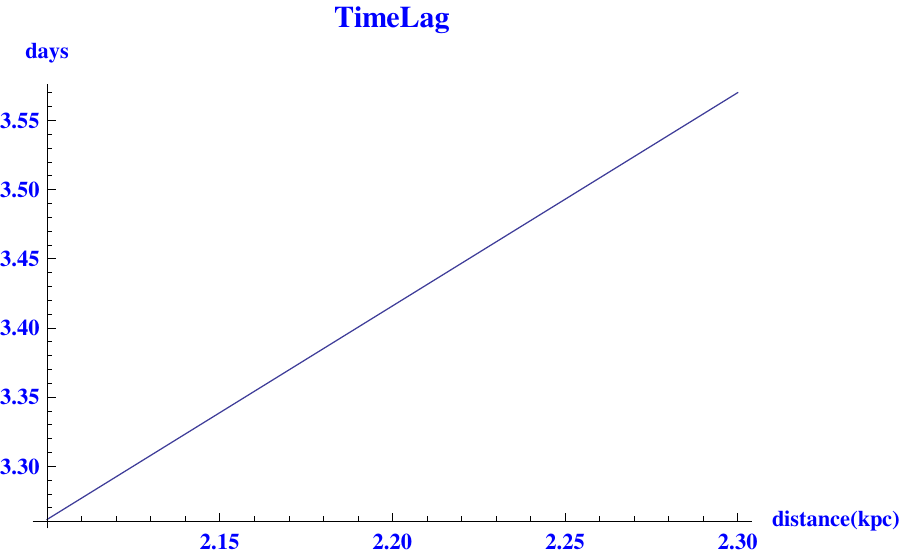}
\caption{The Shapiro delay due to the dark matter potential along the line of sight as a function of distance in the vicinity of the Crab pulsar.}
\label{crab}
\end{figure}

The Crab pulsar (PSR B0531+21) is located at RA=05hr 34m 32 sec and Dec=$22^{\circ} 52.1''$ at a distance of 2.2 Kpc~\cite{atnf}. To calculate the Shapiro delay to this pulsar, we follow the same procedure as used for the delay calculation to PSR~1937B+21~\cite{Desai15}. We provide a brief synopsis of the calculation. More details can be found in Refs.~\cite{Desai,Kahya10,Desai15}. We assume static symmetric geometry and posit the Schwarzschild metric to model the gravitational potential of the dark matter distribution.  The coefficients of the metric are obtained in terms of the density profile and mass distribution by solving Einstein's equations. For the dark matter distribution we use the NFW profile~\cite{NFW} and mass-halo concentration relation from Klypin et al~\cite{Klypin}. With these assumptions, the total Shapiro delay due to the dark matter potential turns out to be 3.4 days for a distance of 2.2~kpc and its variation with distance in the vicinity of the Crab pulsar is shown in Fig.~\ref{crab}. We note that the Shapiro delay for alternate dark matter density profiles  has been calculated in our previous works~\cite{Desai,Kahya10}. Sensitivty to alternate baryonic mass profiles can be found in Ref.~\cite{Desai15}.

To calculate the Shapiro delay from the baryonic matter, we sum the contributions from both the bulge and the disk  and posit  spherical symmetry for the mass distributions of both of them.  We assume a Hernquist profile~\cite{Hernquist} for the mass of the bulge $M_{bulge} = 1.5 \times 10^{10} M_{\odot}$~\cite{Xue}  and Miyamoto and Nagai~\cite{Nagai} profile for the mass of the disk, with the total mass equal to $M_{disk} = 5 \times 10^{10} M_{\odot}$~\cite{Smith}. Therefore, the total mass due to the baryonic component is equal to $6.5 \times 10^{10}  M_{\odot}$. From the total mass, we can calculate the coefficients of the Schwarzschild metric and obtain the total Shapiro delay for both the bulge and the disk. This delay turns out to be 0.12 days for the bulge and 0.32 days for the disk. The total Shapiro delay  to the Crab pulsar after summing all these contributions turns out to be 3.84 days.

Y16 have used the total gravitational potential of the Milky way and with their assumptions, one obtains a Shapiro delay of 
(5.14-15.42) days. Z16 have assumed an NFW profile for the dark matter halo and a Miyamoto-Nagai disk for the baryonic component and have used the Milky way parameters for these from Gomez et al~\cite{Gomez}. From their value of the  gravitational potential, the inferred Shapiro delay is about 1.98 days. Therefore, the delay which we have calculated by integrating the geodesics for the Schwarzschild metric is in-between the values obtained by Y16 and Z16, but is of the same order of magnitude.

We now emphasize some of the key differences between our calculations and those done by Y16 and Z16.
We have calculated the Shapiro delay by solving  the  geodesic deviation equation perturbatively by positing pressureless, static, and spherically symmetric system. To calculate the coefficients of the Schwarzschild metric, we sum the contributions from both the dark matter and the baryonic matter. 
Both Y16 and Z16 have assumed the point mass approximation to calculate the Shapiro delay and used the original formula derived by I. Shapiro in 1964~\cite{Shapiro}, where one gets the logarithm. This is valid only for the case of a first order perturbation theory in the PPN formalism.  For the case of dark matter in addition to the baryonic matter, there are two small parameters: usual 2GM/R$c^2$ and 2$v^2/c^2$, which is due to the constancy of the asymptotic rotation speed of spiral galaxies and these are of the same order. Therefore, the Shapiro delay calculation should be done more carefully  as in this work, and the difference between the two cases can be found in Kahya and Woodard~\cite{Kahya07} and more details in ~Soussa and Woodard~\cite{Soussa}.

Therefore, since our calculation is a fully general relativistic one and does not assume a point source for the gravitating mass and assume a varying density,  our result should be  more accurate than what was done in  Y16 and Z16. Furthermore, Y16 have used the galactic rotation curve from Irrgang et al~\cite{Irrgang} extending upto 10~kpc. to  calculate the  contribution of the Milky way galaxy.  Since the Crab is located at a distance of about 3 Kpc, incorporating  the total potential upto 10 Kpc would result in an overestimate of the Shapiro delay. Moreover, Z16 have not included the contribution due to the bulge. Therefore their calculation would be a slight overestimate of the true value.

\section{Constraints on WEP}
 Once the Shapiro delay for a given mass distribution is calculated along a line of sight to the Crab pulsar, if  photons of different frequencies/energies arrive from the same source  within a time interval ($\Delta t$), after traversing the Cosmos, one can constrain the violations of WEP in terms of the PPN parameter  $\Delta \gamma$  and the calculated Shapiro delay $T_{\rm shapiro}$~\cite{Wei}:
\begin{equation}
\Delta\gamma \leq 2 \frac{\Delta t}{T_{\rm shapiro}}
\end{equation}
We note that $\Delta$t includes contributions from the intrinsic time delay, violation of Lorentz invariance induced delays, delay due to possible non-zero photon mass, and  an additional delay due to cold plasma dispersion along the line of sight (valid only for radio waves)~\cite{grb,Wei}. If we assume that all the other time delays are zero or negligible, we can estimate a limit on $\Delta \gamma$~\cite{Wei}  If we use the nano-second pulse from Crab with flux exceeding 2~MJy~\cite{nanoshot}, which had $\Delta t <0.4$ ns, we get a corresponding limit on $\Delta\gamma < 2.41\times 10^{-15}$. We note that the radio pulse observations include a correction for the dispersion induced delay ($\Delta t_{DM}$). However, an additional assumption made in obtaining a limit on $\Delta \gamma$ from the Crab nano-shot observations is that  the PPN $\gamma$ does not have the same $\frac{1}{\nu}^2$ dependence on frequency as $\Delta t_{DM}$~\cite{Lorimer}. We do however know from extragalactic observations of GRBs  that the PPN $\gamma$ is independent of energy  over a  range of eV to GeV to within $\mathcal{O} 10^{-8}$~\cite{Gao}.

The most precise time difference from multi-frequency monitoring of the Crab pulsar is between the radio and  optical wavelengths equal to $255\pm 21 \mu s$~\cite{Cognard}, where the radio data is obtained from the Nan\c{c}ay radio telescope and optical data from the S-CAM3 imager on the OGS telescope in Tenerife.  
Using this observed value of $\Delta t$, we get $\Delta \gamma<1.54 \times 10^{-9}$. Z16 have also  used other combinations of multi-wavelength observations to set limits on WEP, but these are less stringent than those obtained using radio and optical observations.  Therefore, we only report results on the violation of WEP using radio and optical observations.

\section{Conclusions}

We have calculated the total Shapiro delay to the Crab pulsar due to the gravitational potential along the line of sight.  We included the contributions from both the dark  and baryonic matter. The total delay from the gravitational potential of the dark matter distribution to the Crab pulsar is 3.4 days.  The delay from the baryonic matter is equal to 0.12 days  and 0.32 days for the bulge and disk respectively. Therefore, the total Shapiro delay is equal to 3.84 days. We also reviewed the differences in our calculations of Shapiro delay and those from other groups~\cite{BZhang,YZhang}, which also estimated this delay.  Using this value for Shapiro delay, we then used the same multi-wavelength observations of the Crab pulsar as in  Y16 and Z16~\cite{BZhang,YZhang} to obtain limits on violation of weak equivalence principle  in terms of the PPN parameter $\gamma$. Similar to Y16, we use observations of ``nano-shot'' giant pulses with time-delay $<0.4$~ns  resulting in $\Delta \gamma < 2.41\times 10^{-15}$. We then follow Z16, and use the  time-differences between radio and optical photons from a multi-wavelength observing campaign of the Crab pulsar and obtain $\Delta \gamma < 1.54\times 10^{-9}$.


\acknowledgement
{We thank Richard Woodard for prior collaboration   on the Shapiro delay calculations. E.O.K. acknowledges support from TUBA-GEBIP 2016, the Young Scientists Award Programme.}

\bibliographystyle{spphys}
\bibliography{crab}

\begin{thebibliography}{10}
\providecommand{\url}[1]{{#1}}
\providecommand{\urlprefix}{URL }
\expandafter\ifx\csname urlstyle\endcsname\relax
  \providecommand{\doi}[1]{DOI \discretionary{}{}{}#1}\else
  \providecommand{\doi}{DOI \discretionary{}{}{}\begingroup
  \urlstyle{rm}\Url}\fi

\bibitem{BZhang}
Y.P. {Yang}, B.~{Zhang}, \prd \textbf{94}(10), 101501 (2016).
\newblock \doi{10.1103/PhysRevD.94.101501}

\bibitem{YZhang}
Y.~{Zhang}, B.~{Gong}, \apj \textbf{837}, 134 (2017).
\newblock \doi{10.3847/1538-4357/aa61fb}

\bibitem{Shapiro}
I.I. {Shapiro}, Physical Review Letters \textbf{13}, 789 (1964).
\newblock \doi{10.1103/PhysRevLett.13.789}

\bibitem{Will}
C.M. {Will}, Living Reviews in Relativity \textbf{17}, 4 (2014).
\newblock \doi{10.12942/lrr-2014-4}

\bibitem{Demorest}
P.B. {Demorest}, T.~{Pennucci}, S.M. {Ransom}, M.S.E. {Roberts}, J.W.T.
  {Hessels}, \nat \textbf{467}, 1081 (2010).
\newblock \doi{10.1038/nature09466}

\bibitem{IMB}
R.M. {Bionta}, G.~{Blewitt}, C.B. {Bratton}, D.~{Casper}, A.~{Ciocio}, Physical
  Review Letters \textbf{58}, 1494 (1987).
\newblock \doi{10.1103/PhysRevLett.58.1494}

\bibitem{Kamioka}
K.~{Hirata}, T.~{Kajita}, M.~{Koshiba}, M.~{Nakahata}, Y.~{Oyama}, Physical
  Review Letters \textbf{58}, 1490 (1987).
\newblock \doi{10.1103/PhysRevLett.58.1490}

\bibitem{Longo}
M.J. {Longo}, Physical Review Letters \textbf{60}, 173 (1988).
\newblock \doi{10.1103/PhysRevLett.60.173}

\bibitem{Krauss}
L.M. {Krauss}, S.~{Tremaine}, Physical Review Letters \textbf{60}, 176 (1988).
\newblock \doi{10.1103/PhysRevLett.60.176}

\bibitem{LIGO}
B.P. Abbott, et~al., Phys. Rev. Lett. \textbf{116}(6), 061102 (2016).
\newblock \doi{10.1103/PhysRevLett.116.061102}

\bibitem{Kahya16}
E.O. Kahya, S.~Desai, Phys. Lett. \textbf{B756}, 265 (2016).
\newblock \doi{10.1016/j.physletb.2016.03.033}

\bibitem{Gao}
X.F. {Wu}, H.~{Gao}, J.J. {Wei}, P.~{M{\'e}sz{\'a}ros}, B.~{Zhang}, Z.G. {Dai},
  S.N. {Zhang}, Z.H. {Zhu}, \prd \textbf{94}(2), 024061 (2016).
\newblock \doi{10.1103/PhysRevD.94.024061}

\bibitem{LVCFermi}
B.P. {Abbott}, R.~{Abbott}, T.D. {Abbott}, F.~{Acernese}, K.~{Ackley},
  C.~{Adams}, T.~{Adams}, P.~{Addesso}, R.X. {Adhikari}, V.B. {Adya}, et~al.,
  \apjl \textbf{848}, L13 (2017).
\newblock \doi{10.3847/2041-8213/aa920c}

\bibitem{Meszaros17}
J.J. {Wei}, B.B. {Zhang}, X.F. {Wu}, H.~{Gao}, P.~{M{\'e}sz{\'a}ros},
  B.~{Zhang}, Z.G. {Dai}, S.N. {Zhang}, Z.H. {Zhu}, \jcap \textbf{11}, 035
  (2017).
\newblock \doi{10.1088/1475-7516/2017/11/035}

\bibitem{Boran}
S.~{Boran}, S.~{Desai}, E.~{Kahya}, R.~{Woodard}, ArXiv e-prints  (2017)

\bibitem{Murase}
I.M. {Shoemaker}, K.~{Murase}, ArXiv e-prints  (2017)

\bibitem{Fan}
H.~{Wang}, F.W. {Zhang}, Y.Z. {Wang}, Z.Q. {Shen}, Y.F. {Liang}, X.~{Li}, N.H.
  {Liao}, Z.P. {Jin}, Q.~{Yuan}, Y.C. {Zou}, Y.Z. {Fan}, D.M. {Wei}, \apjl
  \textbf{851}, L18 (2017).
\newblock \doi{10.3847/2041-8213/aa9e08}

\bibitem{Wei}
J.J. {Wei}, H.~{Gao}, X.F. {Wu}, P.~{M{\'e}sz{\'a}ros}, Physical Review Letters
  \textbf{115}(26), 261101 (2015).
\newblock \doi{10.1103/PhysRevLett.115.261101}

\bibitem{Sivaram}
C.~{Sivaram}, Bulletin of the Astronomical Society of India \textbf{27}, 627
  (1999)

\bibitem{damour}
T.~{Damour}, Classical and Quantum Gravity \textbf{29}(18), 184001 (2012).
\newblock \doi{10.1088/0264-9381/29/18/184001}

\bibitem{Kaplan}
S.J. {Tingay}, D.L. {Kaplan}, \apjl \textbf{820}, L31 (2016).
\newblock \doi{10.3847/2041-8205/820/2/L31}

\bibitem{blazar}
J.J. {Wei}, J.S. {Wang}, H.~{Gao}, X.F. {Wu}, \apjl \textbf{818}, L2 (2016).
\newblock \doi{10.3847/2041-8205/818/1/L2}

\bibitem{grb}
H.~{Gao}, X.F. {Wu}, P.~{M{\'e}sz{\'a}ros}, \apj \textbf{810}, 121 (2015).
\newblock \doi{10.1088/0004-637X/810/2/121}

\bibitem{Wei2}
J.J. {Wei}, X.F. {Wu}, H.~{Gao}, P.~{M{\'e}sz{\'a}ros}, \jcap \textbf{8}, 031
  (2016).
\newblock \doi{10.1088/1475-7516/2016/08/031}

\bibitem{Wu17}
X.F. {Wu}, J.J. {Wei}, M.X. {Lan}, H.~{Gao}, Z.G. {Dai}, P.~{M{\'e}sz{\'a}ros},
  \prd \textbf{95}(10), 103004 (2017).
\newblock \doi{10.1103/PhysRevD.95.103004}

\bibitem{nanoshot}
T.H. {Hankins}, J.A. {Eilek}, \apj \textbf{670}, 693 (2007).
\newblock \doi{10.1086/522362}

\bibitem{Desai}
S.~{Desai}, E.O. {Kahya}, R.P. {Woodard}, \prd \textbf{77}(12), 124041 (2008).
\newblock \doi{10.1103/PhysRevD.77.124041}

\bibitem{Kahya10}
E.O. {Kahya}, Physics Letters B \textbf{701}, 291 (2011).
\newblock \doi{10.1016/j.physletb.2011.05.073}

\bibitem{Desai15}
S.~{Desai}, E.O. {Kahya}, Modern Physics Letters A \textbf{31}, 1650083 (2016).
\newblock \doi{10.1142/S0217732316500838}

\bibitem{atnf}
R.N. {Manchester}, G.B. {Hobbs}, A.~{Teoh}, M.~{Hobbs}, \aj \textbf{129}, 1993
  (2005).
\newblock \doi{10.1086/428488}

\bibitem{NFW}
J.F. {Navarro}, C.S. {Frenk}, S.D.M. {White}, \apj \textbf{462}, 563 (1996).
\newblock \doi{10.1086/177173}

\bibitem{Klypin}
A.~{Klypin}, H.~{Zhao}, R.S. {Somerville}, \apj \textbf{573}, 597 (2002).
\newblock \doi{10.1086/340656}

\bibitem{Hernquist}
L.~{Hernquist}, \apj \textbf{356}, 359 (1990).
\newblock \doi{10.1086/168845}

\bibitem{Xue}
X.X. {Xue}, H.W. {Rix}, G.~{Zhao}, P.~{Re Fiorentin}, T.~{Naab},
  M.~{Steinmetz}, F.C. {van den Bosch}, T.C. {Beers}, Y.S. {Lee}, E.F. {Bell},
  C.~{Rockosi}, B.~{Yanny}, H.~{Newberg}, R.~{Wilhelm}, X.~{Kang}, M.C.
  {Smith}, D.P. {Schneider}, \apj \textbf{684}, 1143 (2008).
\newblock \doi{10.1086/589500}

\bibitem{Nagai}
M.~{Miyamoto}, R.~{Nagai}, \pasj \textbf{27}, 533 (1975)

\bibitem{Smith}
M.C. {Smith}, G.R. {Ruchti}, A.~{Helmi}, R.F.G. {Wyse}, J.P. {Fulbright}, K.C.
  {Freeman}, J.F. {Navarro}, G.M. {Seabroke}, M.~{Steinmetz}, M.~{Williams},
  O.~{Bienaym{\'e}}, J.~{Binney}, J.~{Bland-Hawthorn}, W.~{Dehnen}, B.K.
  {Gibson}, G.~{Gilmore}, E.K. {Grebel}, U.~{Munari}, Q.A. {Parker}, R.D.
  {Scholz}, A.~{Siebert}, F.G. {Watson}, T.~{Zwitter}, \mnras \textbf{379}, 755
  (2007).
\newblock \doi{10.1111/j.1365-2966.2007.11964.x}

\bibitem{Gomez}
F.A. {G{\'o}mez}, A.~{Helmi}, A.G.A. {Brown}, Y.S. {Li}, \mnras \textbf{408},
  935 (2010).
\newblock \doi{10.1111/j.1365-2966.2010.17225.x}

\bibitem{Kahya07}
E.O. {Kahya}, R.P. {Woodard}, Physics Letters B \textbf{652}, 213 (2007).
\newblock \doi{10.1016/j.physletb.2007.07.029}

\bibitem{Soussa}
M.E. {Soussa}, R.P. {Woodard}, Classical and Quantum Gravity \textbf{20}, 2737
  (2003).
\newblock \doi{10.1088/0264-9381/20/13/321}

\bibitem{Irrgang}
A.~{Irrgang}, B.~{Wilcox}, E.~{Tucker}, L.~{Schiefelbein}, \aap \textbf{549},
  A137 (2013).
\newblock \doi{10.1051/0004-6361/201220540}

\bibitem{Lorimer}
D.R. {Lorimer}, M.~{Kramer}, \emph{{Handbook of Pulsar Astronomy}} (2004)

\bibitem{Cognard}
T.~{Oosterbroek}, I.~{Cognard}, A.~{Golden}, P.~{Verhoeve}, D.D.E. {Martin},
  C.~{Erd}, R.~{Schulz}, J.A. {St{\"u}we}, A.~{Stankov}, T.~{Ho}, \aap
  \textbf{488}, 271 (2008).
\newblock \doi{10.1051/0004-6361:200809751}

\end{thebibliography}

\end{document}